\documentclass[a4paper,11pt]{article}
\usepackage[dvipsnames]{xcolor}

\usepackage{pos}

\title{Cosmic-ray transport in blazars: diffusive or ballistic propagation?}
 \ShortTitle{Diffusive or ballistic?}

\author*[a,b,c]{Patrick Reichherzer}
\author[a,b]{Julia Becker Tjus}
\author[a,b,d]{Mario Hörbe}
\author[a,b]{Ilja Jaroschewski}
\author[b,e]{Wolfgang Rhode}
\author[a,b]{Marcel Schroller}
\author[c]{Fabian Schüssler}

\affiliation[a]{Ruhr-Universit\"at Bochum, Theoretische Physik IV: Plasma-Astroteilchenphysik, Universit\"atsstrasse 150, 44801 Bochum, Germany}

\affiliation[b]{Ruhr Astroparticle and Plasma Physics Center (RAPP Center), Ruhr-Universit\"at Bochum, 44780 Bochum, Germany}

\affiliation[c]{IRFU, CEA, Université Paris-Saclay, F-91191 Gif-sur-Yvette, France}

\affiliation[d]{University of Oxford, Oxford Astrophysics, Denys Wilkinson Building, Keble Road, Oxford, OX1 3RH, United Kingdom}

\affiliation[e]{Department of Physics, TU Dortmund University, 44221 Dortmund, Germany}

\emailAdd{patrick.reichherzer@ruhr-uni-bochum.de}

\abstract{The detection of a PeV high-energy neutrino of astrophysical origin, observed by the IceCube Collaboration and correlated with a 3$\sigma$ significance with Fermi measurements to the gamma-ray blazar TXS 0506+056, further stimulated the discussion on the production channels of high-energy particles in blazars. Many models also consider a hadronic component that would not only contribute to the emission of electromagnetic radiation in blazars but also lead to the production of secondary high-energy neutrinos and gamma-rays.

Relativistic and compact plasma structures, so-called plasmoids, have been discussed in such flares to be moving along the jet axis. The frequently used assumption in such models that diffusive transport can describe particles in jet plasmoids is investigated in the present contribution. While the transport in the stationary scenario is diffusive for most of the parameter space, a flaring scenario is always accompanied by a non-diffusive phase in the beginning. In this paper, we present those conditions that determine the time scale to reach the diffusion phase as a function of the model parameters in the jet.
We show that the type of the charged-particle transport, diffusive or ballistic, has a large influence on many observables, including the spectral energy distribution of blazars.}
\FullConference{37$^{\rm{th}}$ International Cosmic Ray Conference (ICRC 2021)\\
		July 12th -- 23rd, 2021\\
		Online -- Berlin, Germany}

\begin{document}
\maketitle

\section{Introduction}
%The great advances in real-time multi-messanger astronomy, along with improved resolutions of observatories, has lead to correlations of different messangers to a flaring source. 
%has fueled the discussion about the production channels of those same messangers. 
%For example, the correaltion of the high-energy neutrino observed by IceCube correlates with a 3$\sigma$ significance with Fermi measurements to the gamma-ray blazar TXS 0506+056. These improved observations raise the hurdles for theoretical models to explain the origin of these particles. Promising models are based on relativistically accelerated plasmoids moving with large Lorentz factors along the jet in blazars \cite{2020MNRAS.496.2885H}. 

%\Julia{[das stimmt nur bedingt: wir sagen ja, dass es eine Antikorrelation gibt. Es ist auch eher so, dass man mit diesen Methoden nach Korrelationen sucht und die schlussfolgerung, dass man so diese findet, finde ich misleading, es hat sich eher gezeigt, dass das Bild deutlich komplexer ist; also vielleicht besser:] 
The great advances in real-time multi-messenger astronomy in the past decade has lead to the insight that the flaring behavior of the sources is much more complex than simple, linear correlations. For instance, while there is evidence at the $3\sigma$ level for a correlation between gamma-rays and neutrinos from the blazar TXS~0506+056 in 2017, the same source shows evidence for a longer neutrino flare in 2014/2015, with no enhanced gamma-ray emission. Such an anti-correlation between gamma-rays and neutrinos is observed for several potential neutrino flares as discussed in \citep{2021ApJ...911L..18K}.
%}  

%\Julia{
The temporal variability of blazars indicates that the emission happens in compact, relativistic structures moving along the jet, usually assumed to be plasmoids
%} 
%These plasmoids are typically compact structures that are considered to be spherically symmetric for simplicity, 
with radii that 
%typically cover orders of magnitude between 
on the order of $10^{12}$~m and $10^{14}$~m (e.g.\ \cite{2019NatAs...3...88G, 2020MNRAS.496.2885H}). 
%\Julia{see e.g.\ \textbf{gruppen um Franckowiak, Murase, Winter zitieren, haben wir im Kun-Papier zB auch gemacht}.
Here, the typical approach is to solve the transport equation with a given one-dimensional diffusion coefficient that scales with the energy as $E^{\delta}$ as predicted by quasi-linear theory.
%ANMERKUNG: ES KANN TATSAECHLICH AUCH SEIN; DASS HIER BOHM GENOMMEN WIRD, DAS MUESSTE MAN NOCHMAL NACHSCHAUEN!!.}
Typical correlation lengths of the strong turbulent magnetic field (1 Gauss) %\Julia{hier ist ein Knackpunkt: das Feld ist stark, ja; aber ist es rein turbulent? dann wäre ja auf jeden Fall Bohm richtig; aber es gibt ja auch das helische Feld, gibt es da Hinweise, wie die sich zueinander verhalten was die Stärke betrifft? Die Magnetfeldstärke nimmt mit dem Abstand vom Kern ab, ob sich das Turbulenzlevel auf dem Weg ändert, weiß ich gerade nicht...} contained therein correspond to a few percent of the plasmoid radius. 
From the point of view of high-energy TeV particles with gyration radii of about $10^{10}-10^{12}$~m, these plasmoids appear as compact objects. Due to the extreme turbulent magnetic fields and densities in the plasmoids, which are many times higher than the rest of the jet regions, cosmic-ray acceleration and interaction can
%secondaray particles 
take place in the plasmoids.
In order to properly model the emission signatures from blazars, the acceleration and transport of charged particles in these environments must be understood very well. Here we only focus on the latter by assuming larger time scales of the transport than of the acceleration process. Due to the compact nature of the plasmoids, the correct description of the transport depends on the time that particles can stay in the plasmoids. The obvious simplification of transport as diffusive at all times must be critically questioned and in many cases replaced by a ballistic description. Criteria for the use of the respective adequate treatment of the transport for different particle energies and object properties are derived in the following.

\section{The time-domain}
The time-domain of ejected cosmic rays should be characterized by an initial ballistic regime followed by a transition toward the diffusive regime. The transition time depends on the particle's energies, but also on the magnetic field in which the particle propagates. Typically, for low-energetic particles that reside long enough in an astronomical environment, the diffusive description is appropriate. However, compact objects, high particle energies, or a combination of both, requires a ballistic description of the process, as soon as particles leave the region before the transport becomes diffusive. An alternative case for the need for ballistic propagation arises by treating most recently injected particles that are not yet diffusive, as it is shown in \cite{Recchia2021} for a pulsar.

\begin{figure}
\centering
\includegraphics[width=0.8\textwidth]{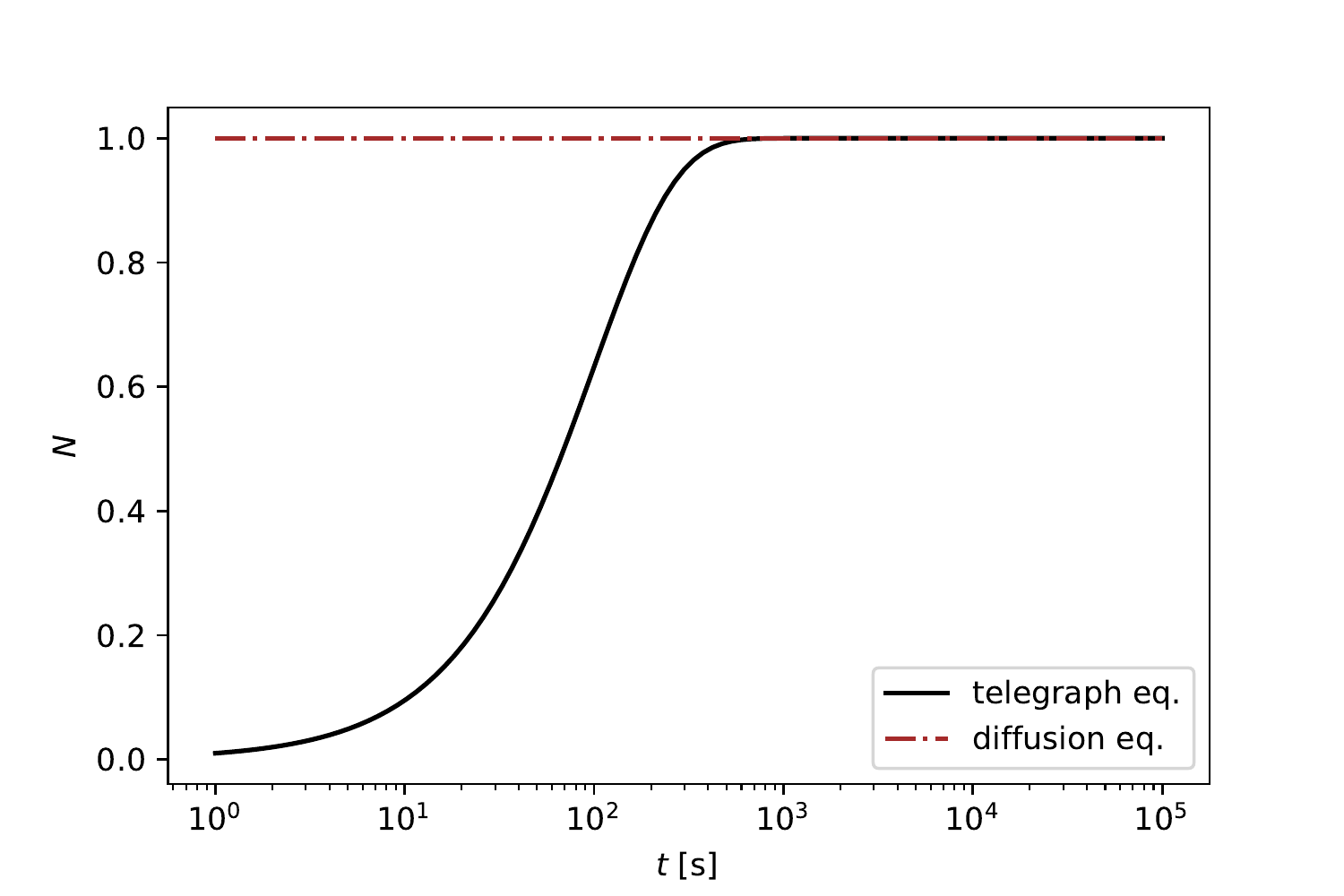}\caption{Comparison of the time evolution of the ratio of particles that are already diffusively propagating on average. The solution of the diffusion equation leads to the fact that particles always propagate diffusively whereas the norm of the solution of the telegraph equation shows an increase in the diffusively propagating particles with time and an approach to the maximum value (see Eq.~(\ref{eq:N})). Here, the time scale is $\tau=100$~s.}\label{fig:1}
\end{figure}

Whereas the description of particle transport via the diffusion equation fails to distinguish between the initial, ballistic propagation and the subsequent diffusive propagation. The telegraph equation has recently been attributed this ability \cite{2013A&A...554A..59L, 2016arXiv160608272T}
\begin{align}
    \frac{\partial f}{\partial t} + \tau \frac{\partial^2f}{\partial t^2} = \sum_i \kappa_i \frac{\partial^2f}{\partial x_i^2}.
\end{align}
Here, the telegraph time scale $\tau$ describes the transition between these two propagation phases. This time scale enables us to make a statement about when the diffusive phase is established and when the description of the particle transport via the diffusion equation is sufficiently accurate. If the initial ballistic phase is neglected, $\tau$ disappears and the telegraph equation turns into the well-known diffusion equation.
The solution of the isotropic telegraph equation yields
\begin{align}
    f_\mathrm{telegraph}(r,t) = \frac{e^{-t/2\tau}}{4\pi\kappa^{3/2}} \left[ \frac{\delta(t-r\sqrt{\tau/\kappa})}{r/\sqrt{\kappa}} I_0 \left( \frac{1}{2}\sqrt{\frac{t^2}{\tau^2}-\frac{r^2}{\kappa \tau}} \right) + \frac{\Theta(t/\sqrt{\tau}-r/\sqrt{\kappa})}{2\tau^{3/2}\sqrt{\frac{t^2}{\tau^2}-\frac{r^2}{\kappa \tau}}} I_1\left( \frac{1}{2}\sqrt{\frac{t^2}{\tau^2}-\frac{r^2}{\kappa \tau}} \right) \right].
\end{align}
Here, $\Theta(...)$ is the Heaviside step function and $I_\nu(...)$ is the modified Bessel function. The norm of the solution of the telegraph equation is unlike the case of the classical diffusion equation time-dependent and yields \cite{2016arXiv160608272T}
\begin{align}\label{eq:N}
    N = 1-\exp\left(-\frac{t_\mathrm{diff,N}}{\tau}\right).
\end{align}
When interpreting the norm as the fraction of particles being diffusive, no particles are diffusive at the beginning, as expected. With increasing time, the number of diffusive particles increases strongly until $N$ approaches 1 for large propagation times. Rearranging the equation leads to a calculation rule for the propagation time required to establish a certain diffusion level
\begin{align}
    t_\mathrm{diff,N} = -\ln{(1-N)}\tau.
\end{align}
This relation is shown Fig. \ref{fig:1} in comparison with the constant number of diffusing particles in the case of modeling the transport with the diffusion equation.

\section{Different transport regimes}
In the previous section, it was shown that particles can only be correctly described by the classical diffusion equation when particles have had long enough to transition to the diffusive phase after the initial ballistic propagation. For isotropic turbulence, it was shown that particles have to travel at least the mean-free path length. By taking into account the dependence of the mean-free path length on the diffusion coefficient and considering the theoretical description of the dependences of the diffusion coefficient on the particle energy and the magnetic field configurations, the required time could be described as a function of the same parameters.\\\\
A key aspect of this equation is the index $\delta$. It is known from numerous theoretical calculations and numerical simulations that particle transport is divided into different regimes with different values for $\delta$. The affiliation of the particles to the different regimes depends mainly on the particle energy and the correlation length of the turbulence. (see \cite{2020MNRAS.498.5051R, 2020PhR...872....1B} for a detailed treatment of the 5 regimes). For particle energies above a few GeV and typical astrophysical environments of interest, such as AGN, the Galaxy, neutron stars, supernovas, etc., only the \textit{resonant-scattering regime} (RSR) and the \textit{quasi-ballistic regime} (QBR), as well as the \textit{transition regime} (TR) between the two, are relevant. The TR extends over a small energy range and therefore will be ignored for simplicity reasons in the following.
It is important to note here that the RSR is also often referred to as a \textit{diffusive regime}. We use the term \textit{diffusive regime}, however, only for the temporal regime in which particles are truly diffusive (see Section 2). This is because regardless of whether the particles belong to one of the energy-dependent transport regimes, particles always propagate ballistically after their injection. 
The transition from the RSR to the QBR is at around $r_g/l_c \approx 1$. 
\begin{figure}
\centering
\includegraphics[width=1.0\textwidth]{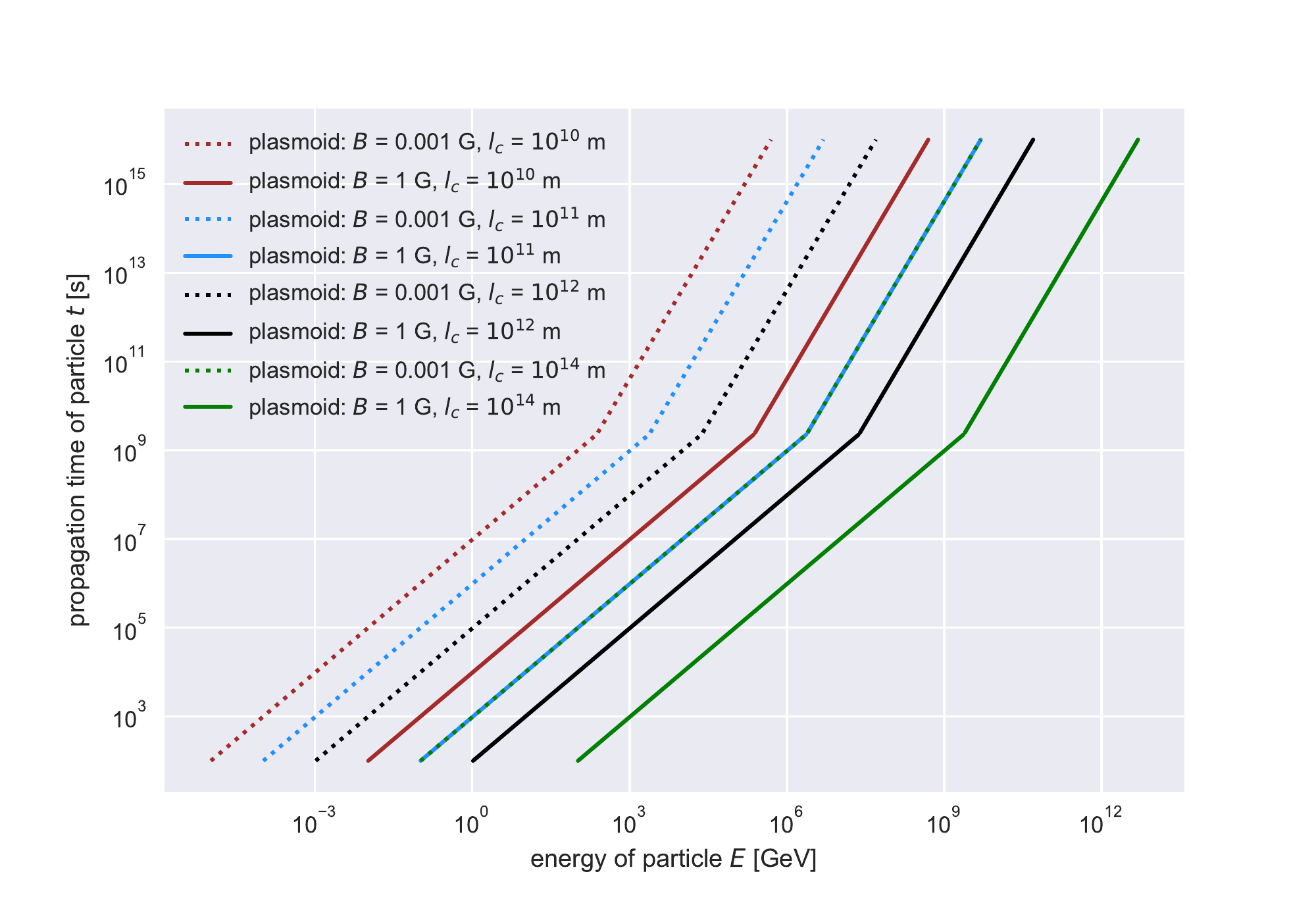}\caption{The lines represent the critical time needed to have $99.9\%$ particles propagating diffusively for the given plasmoid settings. Only for time scales larger than indicated by the lines, a diffusion approach is reasonable.}\label{fig:2}
\end{figure}

The two regimes differ in the point that resonant scattering of charged particles in the QBR contributes little to diffusivity, resulting in $\delta =2$, whereas resonant scattering takes a major part in the diffusion process in the RSR. There is great agreement between theoretical considerations and numerical simulations. For the RSR, however, other effects such as field line random walk contribute to the diffusion coefficient. Moreover, the turbulence level plays an important role for $\delta$ -- QLT predicts $\delta = 1/3$ in the limit of very weak turbulence $\delta B \ll B$, whereas the Bohm diffusion yields $\delta = 1$ for strong turbulence. In between, there must be a transition, which simulations of \cite{2009JGRA..114.1102M} hint for Kraichnan turbulence and \cite{2020MNRAS.498.5051R, reichherzer2021a} find for Kolmogorov-type turbulence. For the limiting case $r_g \ll l_c$, i.e., when the mean-free path length is smaller than the correlation length, field vectors of turbulence would change on larger scales than the particle scattering processes relevant to diffusion occur. An effective background field from the point of view of the particles would emerge that result in a reduced $\delta$ that may be consistent with QLT \cite{2017ApJ...837..140S, 2020PhRvD.102j3016D}. Especially for investigations at these low particle energies, numerical effects on simulation results have to be investigated and minimized \cite{Schlegel2020}.

In summary, the following parameters for the scaling of the diffusion coefficient apply in any case:
\begin{equation}
\delta =  
\begin{cases}
1/3~\mathrm{to}~1&    $for RSR: $r_g \lesssim l_c \\
2&    $for QBR: $r_g \gtrsim l_c
\end{cases}.
\end{equation}
Taking this into account, it becomes evident that higher energy particles take longer to become diffusive. Figure~\ref{fig:2} illustrates this relationship by visualizing equation 1 for different parameter combinations of possible plasmoids moving in the jet. 
The time $t_{\mathrm{diff},N}$ needed for the fraction $N$ of the particles to be diffusive depends on the parameters $E$, $B$ and $l_\mathrm{c}$. Figure \ref{fig:2} shows this condition for different plasmoid parameters. Particles of a given energy have to stay longer in the plasmoid as indicated by the line that corresponds to the plasmoid settings. For particles that leave the plasmoid earlier, a ballistic transport description has to be chosen. Note that the addition of strongly aligned background fields leads to larger parallel diffusion coefficients, which are expressed in a longer initial phase until the diffusive phase is reached. 
\begin{figure}
    \centering
    \includegraphics[width=0.85\textwidth]{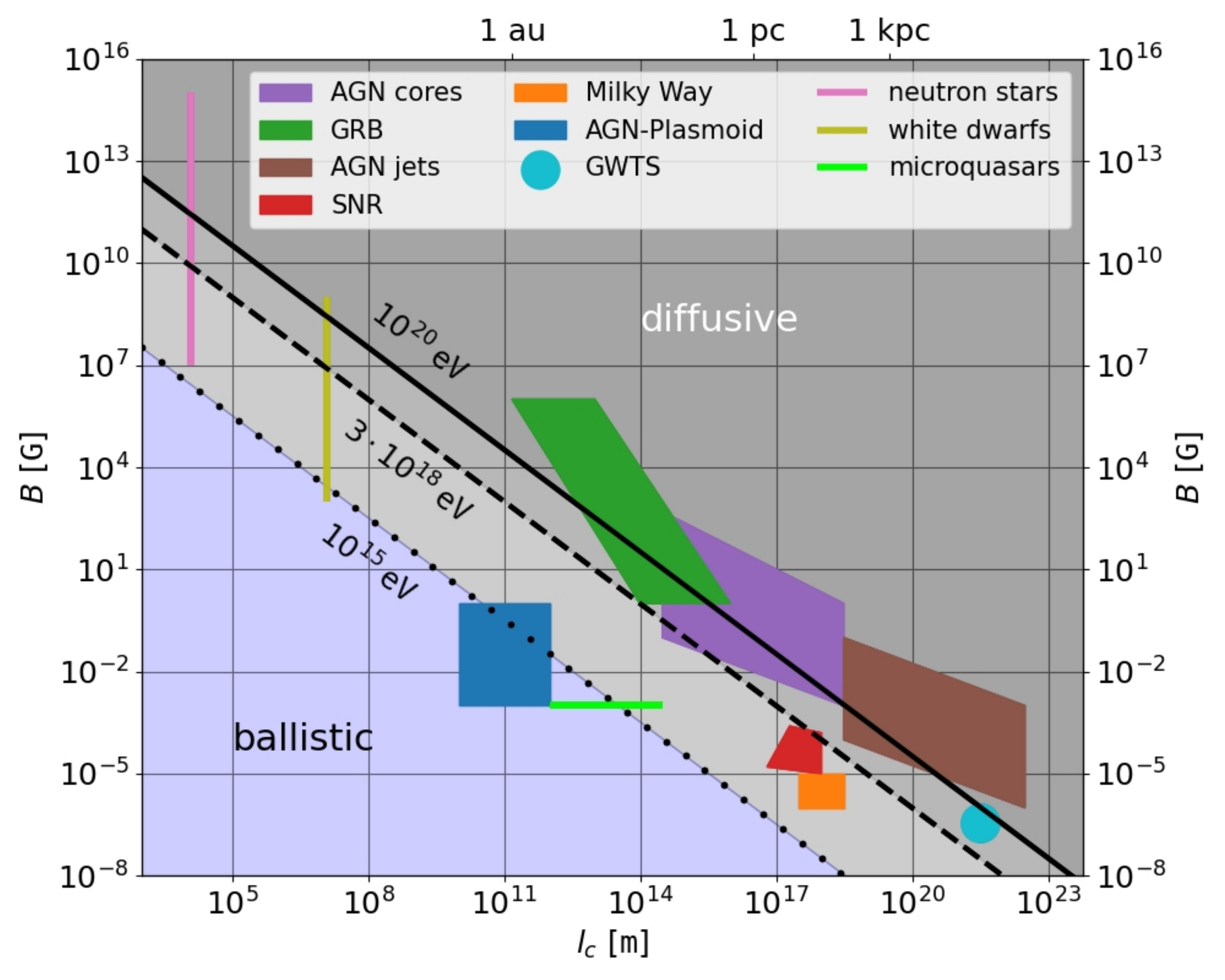}
    \caption{The affiliation of charged particles to the resonant-scattering regime and the quasi-ballisitic regime, depends on the magnetic field strength, the correlation length $l_\mathrm{c}$ of the magnetic field (related to the system size) and the particle energy, where the transition of both regimes is given by $l_\mathrm{c} \approx r_\mathrm{g}$. Under the relationship described in the text between the rapidly growing diffusion coefficient in the QBR and the associated rapid increase in the time required for particles to be diffusive, it is assumed here in a simplifying approach that particles in the QBR regime are described ballistically on the scales of the objects. Consequently, the comparatively small diffusion coefficients in the RSR allow the assumption of a valid diffusive description of the particles. According to this argumentation, the drawn lines of particles of the same energy can shift slightly with a change in the turbulence level and the real extension of the magnetic field.}
    \label{fig:Diagramm_Hillas}
\end{figure}
\section{The space-domain}
The previous analysis has demonstrated that particles in the RSR as well as in the QBR need time to become diffusive. The following simplified picture emerges under consideration of the different values for $\delta$ for RSR and QRB. Particles in QBR will leave usual astronomical sources faster than they have time to become diffusive\footnote{The details must then be calculated on a case-by-case basis according to the equations motivated in the previous section} because of the large diffusion coefficients arising from their fast increase with energy, i.e.~$\kappa \propto E^2$. This picture can now be applied to typical astronomical sources and environments of cosmic rays. Each environment contains different magnetic field configurations covering different magnetic field strengths and coherence lengths of the turbulence, as shown in figure~\ref{fig:Diagramm_Hillas}.
The plot also shows the lines representing the transition from RSR to QBR for particles of the indicated energies. Only particles above the line corresponding to their energy can become diffusive in the typical residence times. 

This plot thus provides a first overview of whether particles of given energy should be diffusively or ballistically propagated in the individual sources. From this observation, it becomes evident that the propagation of high-energy particles in AGN plasmoids only takes place ballistically. Particles with energies of $3\cdot 10^{18}$~eV, for example, must be treated ballistically in AGN plasmoids, whereas propagation in AGN cores takes place diffusively. Only for particle energies below $10^{15}$~eV can a diffusive consideration make sense for certain magnetic field properties in AGN plasmoids.

\section{Outlook}
The considerations presented here can also be extended to estimates of the escape time of particles from specific astrophysical systems. For details, we refer to a follow-up study.
Future discussions and investigations of particle propagation in compact sources must take into account these criteria presented here, and based on them, employ the correct propagation approach -- ballistic or diffusive. In future work, common loss processes for charged particles need to be considered. We defer a more detailed look at acceleration times compared to propagation time to later studies.

%% Full authors list (ONLY FOR COLLABORATIONS)
%\clearpage
%\section*{Full Authors List: \Coll\ Collaboration}
%
%\noindent \textbf{Note comment afterwards:} Collaborations have the possibility to provide an authors list in xml format which will be used while generating the DOI entries making the full authors list searchable in databases like Inspire HEP. For instructions please go to icrc2021.desy.de/proceedings or contact us under icrc2021proc@desy.de.\\
%
%\scriptsize
%\noindent
%first.author$^1$, 
%second.author$^2$, 
%third.author$^3$ % .... more names
%and 
%last.author$^{n}$ \\
%
%\noindent
%$^1$first.affiliation.
%$^2$second.affiliation. % .... more affiliation
%$^{m}$last.affiliation.

\end{document}